\documentclass[]{jfm}

\usepackage{graphicx}
\usepackage{newtxtext}
\usepackage{newtxmath}
\usepackage{mathtools}
\usepackage{booktabs}
\usepackage{natbib}
\usepackage{hyperref}
\hypersetup{
    colorlinks = true,
    urlcolor   = blue,
    citecolor  = black,
}

\newcommand{\RomanNumeralCaps}[1]
\linenumbers

\shorttitle{Quasi-steady aerodynamics of flapping flight}
\shortauthor{O. Pomerenk and L. Ristroph}

\title{Quasi-steady aerodynamics predicts the dynamics of flapping locomotion}

\author{Olivia Pomerenk\aff{1} \and Leif Ristroph\aff{1}\corresp{\email{ristroph@cims.nyu.edu}}}
\affiliation{\aff{1}Courant Institute of Mathematical Sciences, Applied Math Lab, New York University, New York, NY 10012, USA}

\begin{document}

\maketitle

\begin{abstract}

The propulsion of a flapping wing or foil is emblematic of bird flight and fish swimming. Previous studies have identified hallmarks of the propulsive dynamics that have been attributed to unsteady effects such as the formation and shedding of edge vortices and wing-vortex interactions. Here we show that several key features of heaving flight are captured by a quasi-steady aerodynamic model that aims to predict stroke-averaged forces from wing motions without explicitly solving for the flows. We address the forward dynamics induced by up-and-down heaving motions of a thin plate with a nonlinear model which involves lift and drag forces that vary with speed and attack angle. Simulations reproduce the well-known transition for increasing Reynolds number from a stationary state to a propulsive state, where the latter is characterized by a Strouhal number that is conserved across broad ranges of parameters. Parametric, sensitivity, and stability analyses provide physical interpretations for these results and show the importance of accounting for the flow regimes which are demarcated by Reynolds number and angle of attack. These findings extend the phenomena of unsteady locomotion that can be explained by quasi-steady modeling, and they broaden the conditions and parameter ranges over which such models are applicable.

\end{abstract}


\section{Introduction}
Flapping locomotion is widespread among flying and swimming animals such as birds, insects, bats, and fish, all of which propel through fluids using oscillating wings or fins. Given its ubiquity and the potential for biomimetic propulsion systems, there has been significant interest over recent decades in understanding the relevant fluid dynamical mechanisms. Actively propelled flight as well as unsteady motions of bodies passively falling under gravity have been extensively studied through a confluence of experimental, simulatory, and modeling approaches. Laboratory experiments at have involved two-dimensional flow imaging of rigid flapping wings \citep{birch2003influence, vandenberghe2004symmetry}, force sensor measurements of flapping wings \citep{sane2001aerodynamics, sane2002aerodynamic, dickson2004effect, wang2004unsteady}, high-speed camera measurements of the trajectories of falling plates \citep{andersen2005unsteady, huang2013experimental, li2022centre}, and steady-state measurements of the forces on plates at varying angle of attack \citep{li2022centre}, among others. Computational fluid dynamics (CFD) simulations have involved direct numerical solves of the Navier-Stokes equations for free-falling plates in two dimensions \citep{pesavento2004falling, andersen2005unsteady}, and for flapping wings in two dimensions \citep{wang2004unsteady} as well as in three dimensions \citep{sun2002unsteady, pohly2018quasi}. Mathematical modeling efforts have involved the development of quasi-steady models for insect flight \citep{sane2002aerodynamic, birch2003influence, dickson2004effect, ansari2006aerodynamic, pohly2018quasi, wang2016insect} as well as for free-falling plates \citep{andersen2005analysis, andersen2005unsteady, pesavento2006unsteady, hu2014motion, nakata2015cfd, li2022centre, Pomerenk_Ristroph_2025}. These experimental, numerical, and modeling studies have together characterized passive and flapping dynamics at an array of intermediate Reynolds numbers.

A major thrust of this line of research aims to use the fluid dynamical effects identified in experiments and simulations to inform mathematical models of the fluid forces. Quasi-steady models (QSMs) in particular express the instantaneous force in terms of the dynamical state of the wing, i.e. its speed and angle of attack, much in the style of conventional lift-drag aerodynamics. Their advantage is the ease and speed in providing predictions in comparison to experiments and CFD simulations, for which it remains challenging to accurately and efficiently assess a wide range of conditions and parameters \citep{sane2002aerodynamic, pohly2018quasi, wang2016insect}. The disadvantage of QSMs is the loss of accuracy associated with neglecting inherently unsteady effects, which refer to fundamentally time-dependent phenomena associated with the development of the flow field, including vortex generation and shedding and interactions of the wing with its wake flows. These effects are prevalent at the intermediate Reynolds numbers relevant to animal locomotion and for which the combined actions of viscous and inertial flow effects present challenges to modeling \citep{wang2005dissecting, vogel2008modes, klotsa2019above}.

Despite their significant approximations and simplifications, QSMs have demonstrated surprising successes for some problems involving locomotion and motion through fluids. Insect flight is a notable example that defied early attempts to account for the forces during hovering using conventional lift-drag estimates but which was later shown to be well described by more elaborate QSMs that included effects such as delayed stall and rotational lift \citep{sane2003aerodynamics, ansari2006aerodynamic, shyy2010recent, sane2002aerodynamic, liu2005simulation, bergou2010fruit, ristroph2010discovering, ristroph2011paddling, ristroph2013active}. Similarly, the dynamics of the ``falling paper problem'' involving the passive motions of a falling card, including unsteady motions such as fluttering and tumbling as well as steady gliding, have been reproduced by QSMs that include added mass and dynamic center of pressure \citep{andersen2005analysis, andersen2005unsteady, pesavento2006unsteady, wang2012unsteady, huang2013experimental, hu2014motion, nakata2015cfd, li2022centre, Pomerenk_Ristroph_2025}. These successes motivate questions about which other flow-structure interaction phenomena might be brought under the purview of QSMs, and in particular whether flapping-based propulsion is amenable to such a treatment.

Archetypal systems for flapping locomotion involve simple wing shapes actuated with simple kinematics. The up-and-down heaving and plunging motions of a thin, flat, rigid plate have been viewed as emblematic of horizontal propulsion in bird-like flight. For a plate of chord length $\ell$ heaving sinusoidally with frequency $f$ and amplitude $a$ in fluid with density $\rho$ and viscosity $\mu$, the relative strength of inertial to viscous effects can be characterized by the flapping Reynolds number $\Rey_f = \rho a f\ell/\mu$ \citep{vandenberghe2004symmetry, alben2005coherent, lu2006dynamic}. For sufficiently low values of $\Rey_f$, experiments and simulations show that the wing flaps in place and does not progress \citep{vandenberghe2004symmetry,alben2005coherent,lu2006dynamic}. This stationary state gives way to forward locomotion at higher $\Rey_f$ via a symmetry-breaking bifurcation in which the wing accelerates and eventually reaches a nearly steady flight speed. Experiments indicate that the transition occurs at a critical $\Rey^*_f = 20$ to $55$ and is accompanied by hysteresis as well as a region of bistability in which both states can be achieved depending on initial conditions \citep{vandenberghe2004symmetry,vandenberghe2006unidirectional}. Simulations have reported values of $\Rey^*_f$ in the range of $5$ to $50$ and with dependencies on the dimensionless amplitude $a/\ell$ and the solid-fluid density ratio $\rho_s/\rho$ \citep{alben2005coherent,lu2006dynamic}. These works interpret the onset of motion as related to interactions of the wing with the vortex flows generated at its edges during each stroke.

For sufficiently high $\Rey_f$, the terminal flight speed is observed to scale linearly with the flapping kinematic parameters $f$ and $a$ \citep{walker2002functional, vandenberghe2004symmetry, vandenberghe2006unidirectional, alben2005coherent, lu2006dynamic}. This outcome is best quantified by the Strouhal number $St = 2af/U$, which represents the ratio of a characteristic speed of flapping to the mean flight speed $U$. (The numerator is chosen differently in various works, and here we adopt the peak-to-peak amplitude $2a$ as in \citet{alben2005coherent} and elsewhere.) For high $\Rey_f$, the Strouhal number undergoes asymptotic convergence to a constant that has been variously reported as $St = 0.1$ to $0.4$ \citep{wang2000vortex, nudds2004tuning, alben2005coherent, lu2006dynamic, ramananarivo2016flow}. This range is characteristic of swimming and flying organisms, heaving and pitching plates, and flapping foils, and has been shown by experimental and numerical studies to be associated with optimal thrust production during flapping flight \citep{taylor2003flying,nudds2004tuning}. Extensive studies of the accompanying flow fields indicate that the selected value of $St$ is closely associated with the inverted von Kármán wake, which consists of vortices of alternating sign that are deposited in an array due to shedding with each stroke \citep{alben2005coherent, vandenberghe2004symmetry, godoy2008transitions, lu2006dynamic}.

There seems to have been no previous attempt to address the dynamical aspects of flapping propulsion with a QSM. We pursue such here by asking if, despite the clearly documented importance of unsteady effects, the salient phenomena can be reproduced by such a model. If so, what form must it take, and how does this form relate to the previously identified unsteady mechanisms? We focus our investigations on three aspects of the forward dynamics: the state transition from stationary to translating, the timescale defining changes in flight speed from some initial condition, and the speed and Strouhal number characterizing the translating state. We embed our efforts within the same class of models that have been developed for and successfully applied to the falling paper problem. Doing so presents the opportunity to unify these different problems under one framework and to potentially extend the sets of conditions over which they are applicable. Wang and coworkers developed models for the two-dimensional (2D) problem of passive flight of a thin, symmetric plate by accounting for added mass, lift, drag and their associated torques \citep{andersen2005analysis, andersen2005unsteady, pesavento2006unsteady, wang2012unsteady, huang2013experimental, hu2014motion, nakata2015cfd}. Successive iterations have considered wider sets of conditions to drive refinements, including aspects of the center of pressure, the torque model, rotational lift, and the dependencies of the force coefficients on attack angle \citep{huang2013experimental, li2022centre,Pomerenk_Ristroph_2025,hover2004effect}. These developments have generally been done in close connection with CFD simulations and/or experiments and typically in specific contexts that validate the model over limited ranges of Reynolds numbers $\Rey$. A key challenge presented by the flapping propulsion problem is the widely ranging $\Rey = 10$ to $10^5$, where lower values mark the onset of vorticity effects and yet higher values are associated with turbulent boundary layer flows \citep{schlichting2016boundary,anderson2005introduction}. 

In the sections that follow, we first introduce a mathematical model whose main innovations pertain to the aerodynamic coefficients and their dependencies on attack angle and Reynolds number. We then survey the key dynamical outputs for representative conditions and show that the model successfully reproduces the known phenomena. Taking advantage of the computational efficiency of the model and its dimensionless representation, we characterize the dynamical outputs with exhaustive sweeps across all conditions of interest. The results identify a critical criterion for take-off, a Strouhal number for steady forward flight, and a timescale dictating changes in flight speed, all of which are conserved across wide ranges of conditions. A sensitivity analysis then assesses how the model outputs depend on the values of the constant parameters. Finally, we exploit the mathematical tractability of model to analyze the transition via a linear stability analysis and aspects of the emergent forward flight state via an equilibrium analysis. We conclude by interpreting these findings, suggesting further work along this direction, and speculating about applications of the model to related problems.

\section{A quasi-steady model for flapping flight dynamics}

\subsection{Model of forward flight dynamics}

\begin{figure}
\centering
    \includegraphics[width=\textwidth]{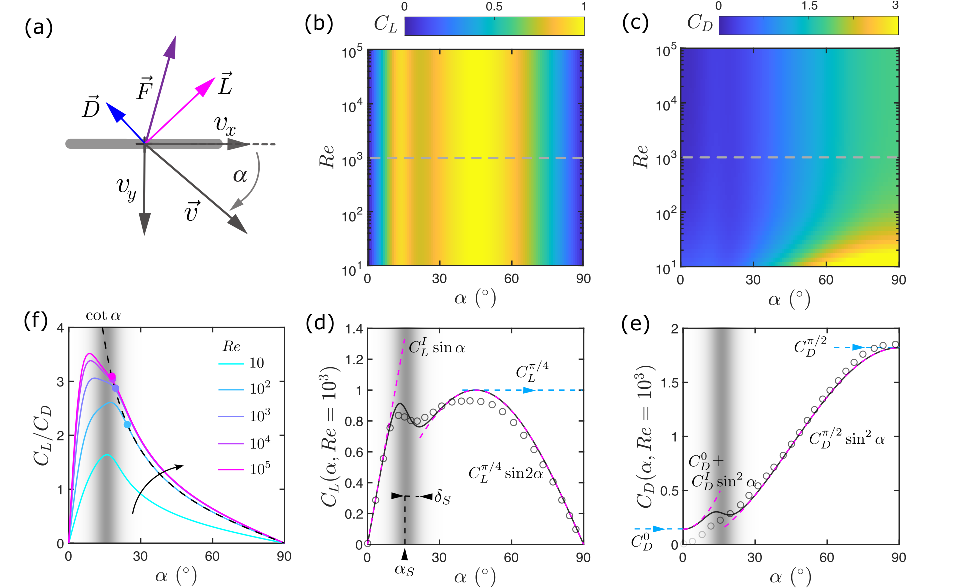}
  \caption{Aerodynamic force coefficients. (a) Definitions of dynamical quantities. Shown here during the downstroke, the plate has horizontal and vertical velocity components $v_x$ and $v_y$ that determine the instantaneous attack angle $\alpha$ and Reynolds number $\Rey$. The total aerodynamic force is resolved into lift and drag. (b) and (c) Color maps of lift and drag coefficients $C_L(\alpha,\Rey)$ and $C_D(\alpha,\Rey)$. (d) and (e) Transects of lift and drag coefficients at fixed $\Rey=10^3$. Dashed magenta curves and the accompanying equations represent the mathematical forms applicable to the attached flow regime at small $\alpha$ and the separated flow regime at high $\alpha$. The grey band shows the transitional region where stall occurs. Dashed blue lines and accompanying labels highlight some parameter values. Open black circles indicate experimental measurements from \citet{li2022centre}. (f) The lift-to-drag ratio $C_L/C_D$ with respect to $\alpha$ at logarithmically-spaced $\Rey$. The graph of $\cot\alpha$ (black dashed curve) and the marked intersections relate to an equilibrium analysis.}
  \label{fig:CL_CD_full}
\end{figure}

We introduce a quasi-steady dynamical model for the driven heaving of a rigid thin plate which is free to move in the direction perpendicular to heaving. To derive such a model, we adopt the form of the quasi-steady model that was developed for the passive free flight of a thin wing \citep{li2022centre,Pomerenk_Ristroph_2025} and which takes the form of the Newton-Euler equations with forces and torques due to gravity, buoyancy, and fluid dynamical effects such as lift, drag, and added mass. The particular case of horizontal motion in a plane and in response to vertical heaving greatly simplifies the aerodynamic considerations by eliminating several degrees of freedom and many of the participating effects. The rotational dynamics need not be evolved because the plate is fixed to lie horizontally in the lab frame as in Fig. \ref{fig:CL_CD_full}(a): $\theta=\omega=0$, where $\theta$ is the plate orientation angle relative to the horizontal and $\omega = \dot{\theta}$ is the angular velocity. This condition simplifies the translational dynamics by eliminating $\omega$-dependent reference frame terms. The vertical motion is also imposed, rather than evolved, and the position is $y(t) = a\sin(2\pi f t)$ and velocity $v_y(t) = \dot{y} = 2\pi a f \cos(2\pi f t)$ for flapping frequency $f$ and amplitude $a$. The horizontal flight motion is therefore the only output, leading to a two-variable dynamical system for $(x,v_x)$ that is nonautonomous due to the time-dependent input $v_y(t)$:
\begin{equation}\label{ODEsystem}
\begin{split}
&\dot{x} = v_{x}\\
&\dot{v}_{x}= \frac{\rho\ell}{2(m+m_{11})}\sqrt{v_x^2+v_y^2} \left[C_L(\alpha,\Rey)v_y - C_D(\alpha,\Rey)v_x\right].
\end{split}
\end{equation}
Here, $x$ and $y$ are lab frame coordinates of the plate center or any other arbitrary reference point on the body. The plate has chord length $\ell$ and mass $m$ per unit length along the span. The fluid has density $\rho$, and the added mass coefficient $m_{11}$ (also per unit span) is associated with edgewise motions of the plate \citep{andersen2005unsteady}.

Notably, the dominant aerodynamic effect that remains is the force associated with translation, which is decomposed into lift and drag components and expressed in the usual high-$\Rey$ form with force coefficients $C_L$ and $C_D$. Figure \ref{fig:CL_CD_full}(a) shows a force diagram at an arbitrary instant. The force coefficients are in general expected to vary with both the attack angle $\alpha=\arctan(v_y/v_x)$ and Reynolds number $\Rey= \rho \ell \sqrt{v_x^2+v_y^2} / \mu$, where $\mu$ is the fluid viscosity coefficient. These quantities vary in time and, due to the involvement of $v_x$, are outcomes of the solution. The mathematical forms for the coefficients $C_L(\alpha,\Rey)$ and $C_D(\alpha,\Rey)$ dictate the behavior of the system and thus whether such a model reproduces the key dynamical features of flapping locomotion. If at all successful, one expects such a model to reproduce forces averaged over times longer than a wing stroke period $1/f$ and hence aspects of the flight dynamics over similarly long timescales. As a quasi-steady treatment, many effects are treated only approximately through the specified force coefficients, and many are neglected altogether. The key assumption that the instantaneous force during unsteady motions is taken as the time-averaged force for the corresponding steady condition (i.e., same speed and attack angle) is not rigorously justified a priori and hence must be assessed by comparing model outputs to known results. Intrinsically unsteady phenomena such as fluctuating forces from vortex shedding are not included \citep{wang2000vortex}.

\subsection{Dimensionless form of the model and its outputs}

Comparison across different parameter values is facilitated by nondimensionalization of Eqs. \ref{ODEsystem}. Following earlier works, we choose characteristic scales for length $\ell$, time $1/f$, and thus speed $\ell f$ \citep{spagnolie2010surprising,alben2005coherent}. Let $v_x' = v_x/(\ell f)$, $x'=x/\ell$, likewise for $v'_y$ and $y'$, and $t'=tf$. Then the imposed driving takes the form $y' = A\sin(2\pi t')$ and $v_y' = \dot{y'}$ for the dimensionless amplitude $A=a/\ell$. The dynamics transform as 
\begin{equation}\label{dimensionless}
    \begin{split}
        \dot{x'} &= v_x' \\
        \dot{v}_x' &= \frac{1}{2M}\sqrt{(v_x')^2 + (v_y')^2}\left[C_L(\alpha,\Rey)v_y'-C_D(\alpha,\Rey)v_x'\right]
    \end{split}
\end{equation}
where $M=(m+m_{11})/\rho \ell^2$ is the dimensionless mass that normalizes the combined solid and added masses (per unit span) by a relevant mass of fluid.

The coefficients $C_L$ and $C_D$ are themselves dimensionless and expressed in terms of dimensionless quantities. The angle of attack $\alpha$ assumes the same form whether dimensional or dimensionless. The dynamical or instantaneous Reynolds number transforms as $\Rey= \rho \ell \sqrt{v_x^2+v_y^2} / \mu = \rho \ell^2 f \sqrt{v_x'^2+v_y'^2} / \mu  = (\Rey_f / A) \sqrt{v_x'^2+v_y'^2}$, where the flapping Reynolds number is $\Rey_f = \rho \ell a f  / \mu$. The system is therefore characterized by three dimensionless input parameters: $M=(m+m_{11})/\rho \ell^2$, $A=a/\ell$, and $\Rey_f = \rho \ell a f  / \mu$. The outputs are the dynamics $(x',v_x') = (x/\ell,v_x/\ell f)$, the results of which will be analyzed to extract summary quantities.

\subsection{Specification of aerodynamic force coefficients}

Lift and drag coefficients for thin rectangular plates have been reported for certain ranges of attack angles and isolated Reynolds numbers \citep{li2022centre, andersen2005analysis, bhati2018role, tomotika1953steady, sane2001control, ehrenstein2013skin}. Based on these partial characterizations, we propose mathematical forms for $C_L(\alpha,\Rey)$ and $C_D(\alpha,\Rey)$ that capture the known flow phenomenology and which involve constant parameters whose values can be informed by reported aerodynamic data. These forms are:

\begin{equation}\label{lift_drag_forms}
\begin{split}
    &C_{L,D}(\alpha,\Rey) = f(\alpha,\Rey) C^{A}_{L,D}(\alpha,\Rey) + \left[1-f(\alpha,\Rey)\right] C^{S}_{L,D}(\alpha,\Rey)\\
    & \quad\quad \textrm{where} \quad f(\alpha,\Rey) = \frac{1}{2}\left[1-\tanh \frac{\alpha-\alpha_S(\Rey)}{\delta_S(\Rey)}\right] \quad \textrm{with} \quad \alpha_S = 15^\circ \quad \delta_S = 5^\circ\\
    &\textrm{Attached flow coefficients:} \\
    & \quad C^A_L(\alpha,\Rey) = C^I_L(\Rey) \sin \alpha \quad \textrm{with} \quad C^I_L = 5 \\
    & \quad C^A_D(\alpha,\Rey) = C^0_D(\Rey) + C^I_D(\Rey) \sin^2 \alpha \quad \textrm{with} \quad  C^I_D = 5\\
    & \quad \quad \quad \textrm{where} \quad C^0_D(\Rey) = C^{0,FT}_D + \frac{C^{0,SF}_D}{\sqrt{\Rey}} \quad \textrm{with} \quad  C^{0,FT}_D = 0.1 \quad C^{0,SF}_D = 1.3 \\
    &\textrm{Separated flow coefficients:} \\
    & \quad C^S_L(\Rey) = C^{\pi/4}_L(\Rey) \sin 2\alpha \quad \textrm{with} \quad C^{\pi/4}_L = 1 \\
    & \quad C^S_D(\Rey) = C^{\pi/2}_D(\Rey) \sin^2 \alpha\\
    &\quad\quad\textrm{where} \quad C^{\pi/2}_D(\Rey) = C^{\pi/2,HR}_D+\frac{C^{\pi/2,LR}_D}{\Rey} \quad \textrm{with} \quad C^{\pi/2,HR}_D=1.8 \quad  C^{\pi/2,LR}_D=20.\\
\end{split}
\end{equation}

These equations represent surfaces $C_L(\alpha,\Rey)$ and $C_D(\alpha,\Rey)$ that are plotted in Figs. \ref{fig:CL_CD_full}(b) and (c). The attack angle $\alpha\in[0,\pi/2]=[0,90^\circ]$ is shown over the conventional aerodynamic range, and the coefficients are readily extended over all possible orientations $\alpha\in[0,2\pi)$ by consideration of the dual symmetries of the plate \citep{li2022centre}. The displayed range of the Reynolds number $\Rey \in [10,10^5]$ reflects the intermediate regime over which the model can be expected to apply. The lift map is strictly uniform with respect to $\Rey$, and the drag map is largely so except for notable variations at lower values. The forces vary with $\alpha$ in ways that are more clearly seen in the transects of Figs. \ref{fig:CL_CD_full}(d) and (e) at a representative $\Rey=10^3$, and these panels graphically interpret the terms and parameters in the model equations. Comparison to the experimental measurements (open circles) of \citet{li2022centre} shows good agreement in the forms of the curves while not faithfully reproducing all features of the data. Note that these experiments reflect only pressure-based or normal forces and hence the measured $C_D$ at low angles $\alpha<\alpha_S$ (faded data points) are underestimates that neglect skin friction as a force tangential to the plate \citep{li2022centre}. Additional checks against experimental and computational studies at $Re=10^2$ \citep{wang2004unsteady} and $Re=10^5$ \citep{ananda2015measured} indicate good agreement for low angles and generally at the level of trends in the curves and rough magnitudes.

The equations (\ref{lift_drag_forms}) above are structured to reflect the program we have used for specifying the coefficient formulas. First, we assume two distinct aerodynamic regimes which are respectively associated with attached flow for low $\alpha$ and fully separated flow at high $\alpha$. The function $f(\alpha,\Rey)$ serves the role of selecting the regime. It is the sigmoid-shaped logistic curve that takes on values near 1 for low $\alpha$, drops near $\alpha_S$, and approaches 0 for high $\alpha$. The critical angle $\alpha_S$ marks the stall transition, and the factor $\delta_S$ specifies the angular width or range over which the transition occurs. Both may in principle depend on $\Rey$, but available data at $\Rey=10^3$ and $10^5$ show stall to occur over the range $\alpha = 10^\circ$ to $20^\circ$, hence our choices for the constant values $\alpha_S = 15^\circ$ and $\delta_S = 5^\circ$ \citep{li2022centre, ananda2015measured}. The transitional region is marked by grey stripes in Figs. \ref{fig:CL_CD_full}(d) and (e).

Next we set about specifying the coefficients $C^{A}_{L,D}(\alpha,Re)$ for the attached flow regime, focusing first on the dependence on $\alpha$ then $\Rey$. The lift coefficient $C^A_L \propto \sin \alpha$ follows the form of Kutta-Joukowski theory for thin wings \citep{anderson2011ebook} and is consistent with measurements at various $\Rey$ showing approximately linear increase in lift with attack angle for low $\alpha$ \citep{li2022centre, ananda2015measured}. The prefactor $C^I_L(\Rey)$ is the slope or inclination of the lift curve at small $\alpha$, and it may in principle depend on $\Rey$. Its value is $2\pi$ in potential flow theory \citep{anderson2011ebook, gutierrez2021lift}, and experimental studies across widely ranging $\Rey =10^2$ to $10^5$ report somewhat lower values between 4 to 6 with no systematic dependence on $\Rey$ \citep{ananda2015measured, li2022centre, gutierrez2021lift}. Hence we choose the constant $C^I_L(\Rey)=5$ as a representative value across all $\Rey$. The drag model $C^{A}_{D}(\alpha,\Rey)$ has two terms, the first of which represents the fluid resistance present for edgewise motions at $\alpha=0$. It involves the constant term that represents the (small) pressure drag associated with the finite thickness of the plate \citep{andersen2005unsteady, anderson2005introduction}. It is expected to be of the same order as the thickness-to-chord ratio, and the value $C^{0,FT}_D = 0.1$ is chosen as typical of wing-plates considered in previous studies \citep{taylor2003flying,vandenberghe2004symmetry,andersen2005analysis,alben2005coherent,li2022centre}. The second term captures the effect of skin friction, and the prefactor value $C^{0,SF}_D = 1.3$ and scaling as $1/\sqrt{\Rey}$ are the classical results of the Blasius boundary-layer theory calculation \citep{bhati2018role, tomotika1953steady}. The second term in $C_D^A(\alpha,\Rey)$ represents lift-induced drag, which arises due to the change in effective attack angle due to lift generation \citep{anderson2011ebook, anderson2005introduction}. Its form as $\sin^2 \alpha$ reflects the quadratic dependence on the lift coefficient according to Prandtl's lifting-line theory \citep{anderson2005introduction, gutierrez2021lift}, which also predicts the constant prefactor to depend on the span-to-chord aspect ratio. The chosen value $C_D^I=5$ was determined empirically in \citet{li2022centre} for plates with aspect ratios of 5 to 10.

We carry out similar procedures for the coefficients $C^{S}_{L,D}(\alpha,\Rey)$ in the separated flow regime applicable to high angles beyond the stall transition. Considering lift, the dependence on $\alpha$ of $C^{S}_{L}\propto \sin 2\alpha$ has been adopted in previous modeling work and reflects the observation that lift tends to be maximal near $\alpha = \pi/4 = 45^\circ$ and fall off toward lower and higher values, as expected by symmetry \citep{li2022centre, andersen2005unsteady}. The maximal value of the lift coefficient has been reported to be 0.7 to 1.3 in studies across widely ranging $\Rey$, motivating our choice of the constant $C_L^{\pi/4}=1$ \citep{li2022centre, andersen2005unsteady, ananda2015measured}. Considering drag, the dependence $C^{S}_{D} \propto \sin^2 \alpha$ has similarly been used previously \citep{li2022centre, andersen2005unsteady} and models the observed monotonic increase up to $\alpha = \pi/2 = 90^\circ$. The maximal value of $C^{\pi/2}_D(\Rey)$ is the drag coefficient for a plate in normal flow, which for bluff bodies is constant at higher $\Rey$ and scales as $1/\Rey$ at low $\Rey$ \citep{jones1961drag,bhati2018role}. The associated values $C^{\pi/2,HR}=1.8$ and $C^{\pi/2,LR}=20$ are chosen as fits to measurements across $\Rey \in (10,10^8)$ \citep{bhati2018role}.

In summary, the lift and drag coefficients $C_{L,D}(\alpha,\Rey)$ have functional dependencies on attack angle and Reynolds number that are informed by aerodynamic theory. They involve 9 constants whose values are informed by previous measurements: $\alpha_S = 15^\circ$, $\delta_S = 5^\circ$, $C^I_L = 5$, $C^{0,FT}_D = 0.1$, $C_D^{0,SF} = 1.3$, $C^I_D = 5$, $C^{\pi/4}_L = 1$, $C^{\pi/2,HR}_D = 1.8$, $C^{\pi/2,LR}_D = 20$. The first two define the stall transition, the following four define the attached flow regime for small $\alpha$, and the last three define the separated flow regime for large $\alpha$.

\section{Characterization of flight dynamical behaviors}

Here we characterize the dynamical behaviors resulting from numerical solutions of the proposed quasi-steady model and thereby demonstrate that it closely reproduces the key characteristics of flapping-based propulsion. To this end, Eqs. \ref{dimensionless} are integrated numerically in time via MATLAB's built-in solver \textit{ode15s} that is optimized for systems of stiff ordinary differential equations. Its algorithm is based on numerical differentiation formulas up to fifth order, and variable time-stepping ensures a specified degree of accuracy. Convergence tests determine the values of the relative $\textrm{RelTol}=10^{-10}$ and absolute $\textrm{AbsTol}=10^{-9}$ tolerances used for all results presented here. The initial condition for the translational speed is $v_x'(0)=0.01$, which represents a small perturbation whose purpose is to break the left-right symmetry. We choose $M=A=1$ for simplicity here, and the effects of varying these parameters will be systematically assessed in the next section. 

 \begin{figure}
\centering
    \includegraphics[width=\textwidth]{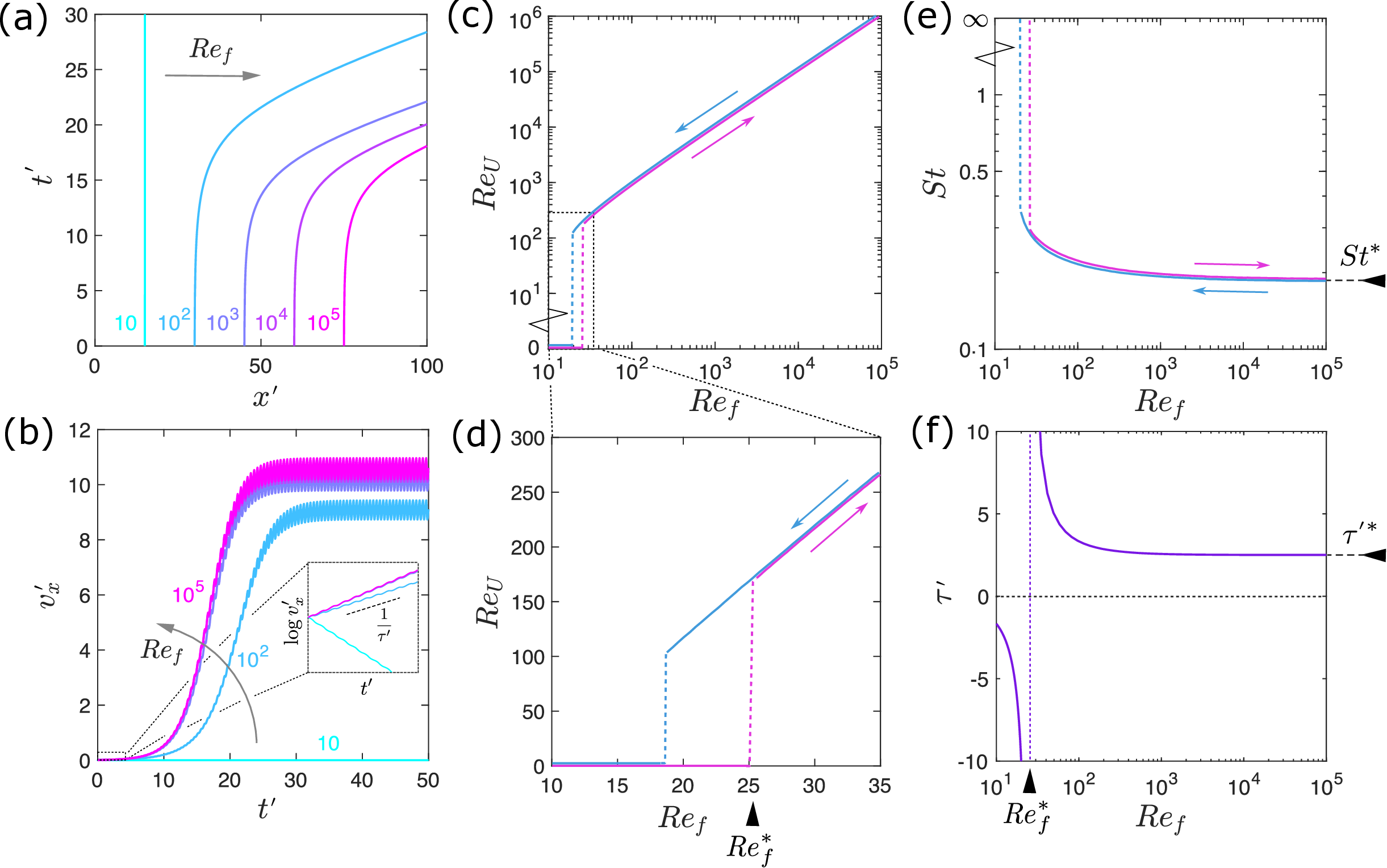}
   \caption{Model dynamics reproduce characteristic flapping flight behaviors. (a) Dimensionless position $x'$ versus $t'$ for the indicated values of $\Rey_f$. (b) Dimensionless horizontal speed against time. Inset: Log-linear plot for early times. (c) Terminal flight speed versus flapping frequency, shown in dimensionless forms as the horizontal Reynolds number $\Rey_{U}$ against $\Rey_f$. Low $\Rey_f$ leads to a stationary state, while higher $\Rey_f$ yields forward flight following a linear relation $\Rey_{U} \propto \Rey_f$. (d) Magnified view for low $\Rey_f$ showing hysteresis and bistability at the transition to forward flight. Stability of the stationary state is lost at the critical value $\Rey_f^*=25$. (e) Strouhal number $St$ plotted against $\Rey_f$, with the value $St^* = 0.2$ characterizing $\Rey_f\gg\Rey_f^*$. (f) Dimensionless exponential timescale $\tau'$ plotted against $\Rey_f$, where the sign of $\tau'$ indicates decay $(-)$ or growth $(+)$ of speed and hence stability or instability of the stationary state. The change of sign indicates the state transition at $\Rey_f^*=25$, and the limiting value $\tau'^*=2.5$ characterizes $\Rey_f\gg\Rey_f^*$.}
  \label{fig:bifurcation}
\end{figure}

Figure \ref{fig:bifurcation}(a) displays curves of the dimensionless displacement $x' = x/\ell$ over time for logarithmically increasing $\Rey_f=10,10^2,\dots,10^5$, where each curve is associated with a different fixed $\Rey_f$ that increases from left to right. The initial positions are spaced apart for ease of visibility. For small $\Rey_f=10$ (left vertical line, cyan), the plate remains stationary for all time. In contrast, higher $\Rey_f \geq 100$ (right curves, darker cyan and magenta) induces the plate to take off into forward flight. Figure \ref{fig:bifurcation}(b) presents these same simulation results in terms of dimensionless speeds $v_x'$ over time for the same choices of $\Rey_f$. Evidently, flapping of a plate induces one of two possible terminal or long-time behaviors determined by the value of the flapping Reynolds number. For low  $\Rey_f$, the plate assumes a stationary state in which it flaps in place with no forward progression. For moderate to high $\Rey_f$, it reaches a forward flight state characterized by constant stroke-averaged translational speed with relatively small fluctuations within each stroke. The inset of Fig. \ref{fig:bifurcation}(b) is a rescaled log-linear plot of the early-time data, from which it is evident that the speed either decays (negatively sloped cyan curve, $\Rey_f = 10$) or grows (positively sloped curves, $\Rey_f = 10^2$ to $10^5$) exponentially from its initial value.

Further details regarding the terminal states and the transition are provided in Fig. \ref{fig:bifurcation}(c) and the magnified view of (d). Here the mean speed $U = \overline{v}_x = \overline{v}_x' \ell f$ during the terminal state is used to compute the forward flight Reynolds number $\Rey_{U}=\rho U\ell/\mu$ based on the horizontal motion. Panel (c) reports the output motion $\Rey_U$ versus the input flapping Reynolds number $\Rey_f$, revealing a linear relationship for sufficiently large $\Rey_f$. The more complex behavior at low $\Rey_f$ is clarified by the zoomed-in view in (d), which reveals a hysteresis loop that brackets a range of bistability in which the stationary and forward flight states coexist. This structure is revealed by a procedure in which we incrementally increase $\Rey_f$ in a stepwise fashion, where each step is initialized with the previous step's flight speed and $\Rey_f$ is held constant for sufficiently long that the forward dynamics equilibrate and $\Rey_U$ can be extracted. The results of this upward sweep are shown as the magenta curve, and it is followed by a downward sweep shown in cyan. These data identify a lower value $\Rey_f = 18$ below which only the stationary state is observed and a higher value of $25$ above which only the forward flight state is seen. As summarized in Table \ref{tab:comparison_table}, these results can be compared to the hysteretic dynamics reported in previous experiments to occur around $\Rey_f = 20$ to 55 \citep{vandenberghe2004symmetry}, as well as the more complex transition to locomotion seen in numerical simulations around $\Rey_f = 15$ to $50$ \citep{alben2005coherent}.  

These same data are recast in Fig. \ref{fig:bifurcation}(e) in terms of the Strouhal number $St = 2af/U = 2\Rey_f/\Rey_U$ attained across varying $\Rey_f$. Beyond the onset to locomotion, $St$ rapidly drops to about 0.3 and thereafter changes little as it asymptotically approaches a constant value near $St^* = 0.2$ for large $\Rey_f$, where $St^* = St(\Rey_f \gg \Rey_f^*)$. As summarized in Table \ref{tab:comparison_table}, these results can be compared to the range $St = 0.15$ to 0.4 that has been reported to apply to the cruising flight and steady swimming of many diverse animals and which has been associated with efficient flapping locomotion \citep{taylor2003flying, nudds2004tuning, alben2005coherent}.  

\begin{table}
\centering
\begin{tabular}{p{2cm}p{3cm}p{4cm}p{3.5cm}}
Quantity & Current study & Previous study & Reference  \\
\midrule
$\Rey_f^*$ & 25 & 20 to 55 \newline 15 to 50 & \citet{vandenberghe2004symmetry} \newline \citet{alben2005coherent} \\
\midrule
$St^*$ & 0.2 & 0.2 to 0.4 \newline 0.21 to 0.25 \newline 0.15 to 0.2 & \citet{taylor2003flying} \newline \citet{nudds2004tuning} \newline \citet{alben2005coherent} \\
\bottomrule
\end{tabular}
\caption{Comparison of predictions from the current study for the quantities $\Rey_f^*$ and $St^*$ against previously reported values from experiments and direct numerical simulations.}\label{tab:comparison_table}
\end{table}

Lastly, we characterize the timescale associated with the observed exponential decay or growth of the flight speed at early times. Building on the analysis shown in the inset of Fig. \ref{fig:bifurcation}(b), we fit lines to the log-linear data of $v_x'(t')$ across values of $\Rey_f$ and associate the inverse of the slope with the decay/growth timescale $\tau'$. Negative values $\tau'<0$ are associated with decay and hence stability of the stationary state, and positive $\tau'>0$ with growth and thus instability that drives the body into forward flight. The extracted data $\tau'(\Rey_f)$ are shown as the curve in Fig. \ref{fig:bifurcation}(f). These results identify the critical value $\Rey_f^*=25$ (vertical dotted line) below which $\tau'<0$ and above which $\tau'>0$. This value matches to the upper bound for the hysteretic region as detailed in Fig. \ref{fig:bifurcation}(d), and this correspondence is understood by noting that both results pertain to the loss of stability of the stationary state. These findings indicate that the sign change of $\tau'$ can be used generally to identify the transitional value $\Rey_f^*$ that is shown here to be 25 for $M=A=1$ and whose characterization across the parameter space is taken up in the next section. Further, for large $\Rey_f \gg \Rey_f^*$ the timescale asymptotically approaches a constant value $\tau'^*=\tau'(\Rey_f \gg \Rey_f^*)$, and its parametric dependence will be assessed in what follows. We are not aware of previous studies characterizing this timescale, and hence the results presented here are predictions that may be compared to the results of future experiments and simulations.

\section{Characterizations across physical and kinematic parameters}

\begin{figure}
\centering
    \includegraphics[width=\textwidth]{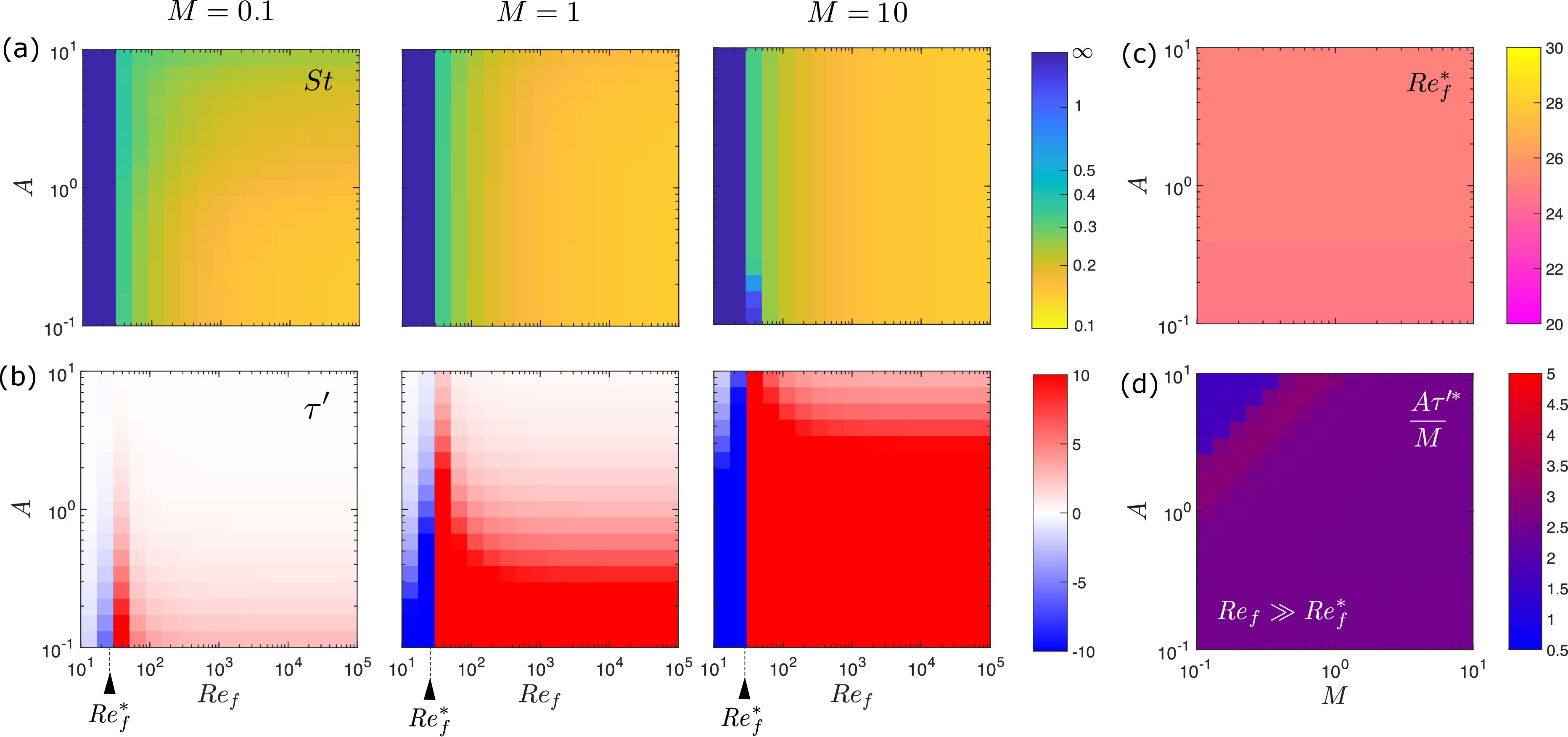}
   \caption{Tableau characterizing the dependencies of the key output quantities on the input parameters $M$, $A$ and $\Rey_f$. Each pixel in a given map corresponds to a run of the simulation at the indicated parameter values. (a) and (b) The Strouhal number $St$ and characteristic growth/decay time $\tau'$. The former is nearly constant with value $St^*=0.2$ for large $\Rey_f$ and across all $M$ and $A$. (c) The critical value $\Rey_f^*$, defined as the point at which $\tau'$ changes sign. This is a nearly constant field in $M$ and $A$ with $\Rey_f^*=25$. (d) The rescaled growth/decay timescale $A\tau'^*/M$ for fixed $\Rey_f=10^5\gg\Rey_f^*$ displays a nearly uniform map implying that $\tau'^* \approx 2.5 M/A$.}
  \label{fig:parameter_sweep}
\end{figure}

Having confirmed the model's ability to reproduce the key phenomena for the intermediate values $M=A=1$, we next conduct a sweep throughout the full space of the physical and kinematic input parameters $M$, $A$, and $\Rey_f$. The output quantities of interest include: the terminal Strouhal number $St$ attained, the characteristic growth or decay timescale $\tau'$, and the critical bifurcation parameter $\Rey_f^*$ that marks the transition from the stationary state to forward flight. It will also be of interest to consider the high-$\Rey_f$ asymptotic values of the former two parameters: $St^* = St(\Rey_f \gg \Rey_f^*)$ and $\tau'^* = \tau'(\Rey_f \gg \Rey_f^*)$. The results are compiled in Fig. \ref{fig:parameter_sweep}. The explored values $M\in\lbrace 0.1,1,10\rbrace $, $A\in[0.1,10]$, and $\Rey_f\in[10,10^5]$ are chosen to span the wide-ranging parameters applicable to flapping-based swimming and flight of animals \citep{pennycuick1996wingbeat,taylor2003flying,vandenberghe2004symmetry,fish2016hydrodynamic,kang2018experimental}.

Displayed in the panels of Fig. \ref{fig:parameter_sweep}(a) are terminal values of $St$, where each panel scans the space of $(\Rey_f,A)$ and the three panels correspond to $M = 0.1$, 1 and 10. For lower values of $\Rey_f$ below the transition, the stationary state prevails and $St \rightarrow \infty$ (dark blue) across all values of $M$ and $A$. For higher values of $\Rey_f$ beyond the transition, the Strouhal number abruptly reaches finite values and saturates to the selected value of about $St^* = 0.2$ (yellow-orange). In this regime, $St$ shows only minor variations with $M$ and $A$. These results seem to corroborate the view that forward flapping flight dynamics is universal in the sense that this key metric is conserved across the vast majority of the parameter space \citep{taylor2003flying, nudds2004tuning, alben2005coherent, ramananarivo2016flow}. 

Displayed in the panels of Fig. \ref{fig:parameter_sweep}(b) are the dimensionless growth/decay times $\tau'$. Across all $M$ and $A$, it is seen that $\tau'<0$ (blue tones) for sufficiently low $\Rey_f < \Rey_f^*$ and $\tau'>0$ (red tones) for higher $\Rey_f > \Rey_f^*$. Thus, the existence of the transition from the stationary state to locomotion is maintained for all conditions, and the transitional value $\Rey_f^*$ itself appears similarly preserved. To assess this quantitatively and systematically across the space, we extract $\Rey_f^*(M,A)$ via the change of sign in $\tau'$ and compile the results in the map of Fig. \ref{fig:parameter_sweep}(c). The panel is notably uniform in color, indicating that the value of the critical flapping Reynolds number is highly conserved near $\Rey_f^* = 25$. Additional investigations indicate that the value $\Rey_f=18$ marking the lower boundary of the hysteresis window is also conserved across $M$ and $A$.  

Further, comparisons within each panel of Fig. \ref{fig:parameter_sweep}(b) show that $|\tau'|$ generally decreases (lighter tones) with $A$, and comparisons across the panels show that $|\tau'|$ increases (darker tones) with $M$. This suggests a relationship of the form $\tau'(M,A) \sim M/A$, and indeed rescaling as $A\tau'/M$ proves to largely remove the variations, as shown by the map of Fig. \ref{fig:parameter_sweep}(d). Here, we fix $\Rey_f  = 10^5$ as representative of $\Rey_f \gg \Rey_f^*$, and it is seen that $A\tau'^*/M$ falls within a narrow range of 1 to 3 as $M$ and $A$ are each varied over two orders of magnitude. 

In summary, these parametric sweeps reveal three quantities that are highly conserved: 1) the selected Strouhal number $St^* = 0.2$ holds for sufficiently high $\Rey_f>\Rey_f^*$ and across all $M$ and $A$; 2) the rescaled dynamical timescale $A\tau'^*/M = 2\pm1$ similarly holds; and 3) the critical Reynolds number $\Rey_f^*=25$ holds for all $M$ and $A$. The significance of these results is that they support the view that some key dynamical features are common to flapping-based propulsion across wide ranges of physical conditions, parameter values, and scales. Such universality has been recognized previously with respect to the Strouhal number and to some degree the critical Reynolds number, and future studies may now investigate the newly identified timescale.

\section{Sensitivity analysis of model parameters}

We next conduct a sensitivity analysis of the constant parameters appearing in the aerodynamic model in order to determine their influence on the dynamical outputs. Specifically, the full slate of parameters include the 9 constants $C_L^I$, $C_D^{0,FT}$, $C_D^{0,SF}$, $C_D^I$, $\alpha_S$, $\delta_S$, $C_L^{\pi/4}$, $C_D^{\pi/2,HR}$, and $C_D^{\pi/2,LR}$, the first 4 of which pertain to the attached flow regime, the next 2 to the stall transition, and the last 3 to the separated flow regime. The dynamical outputs of interest are $\Rey_f^*$, $\tau'^*$, and $St^*$. For the purpose of this analysis, we consider fixed values $M=A=1$, which should be viewed as generally representative given that the outputs are highly conserved. Similarly, we consider fixed $\Rey_f=10^5$ as representative of conditions $\Rey_f \gg \Rey_f^*$ where the asymptotic values $St^*$ and $\tau'^*$ are attained. We define a dimensionless sensitivity score to each output quantity $Q$ with respect to each input parameter $p$:
\begin{equation}
    S_{Q,p} = \left| \frac{\partial Q}{\partial p}\frac{p_0}{Q_0}\right| \approx \left|\frac{\Delta Q/Q_0}{\Delta p / p_0}\right|.
\end{equation}
Here $p_0$ is nominal parameter value and $Q_0 = Q(p_0)$ is the associated nominal value of the output quantity extracted from the numerical solution. The difference approximation to the derivative allows the score to be estimated by running computations for a perturbed parameter value $p = p_0 + \Delta p$ and extracting the change $\Delta Q = Q - Q_0$ in the output quantity. We implement a two-sided approximation by averaging the results of perturbations of size $\pm10\%$ in each parameter, i.e. $\Delta p = \pm 0.1 p_0$.

\begin{figure}
\centering
    \includegraphics[trim={0cm 6.7cm 0cm 15cm},clip,width=\textwidth]{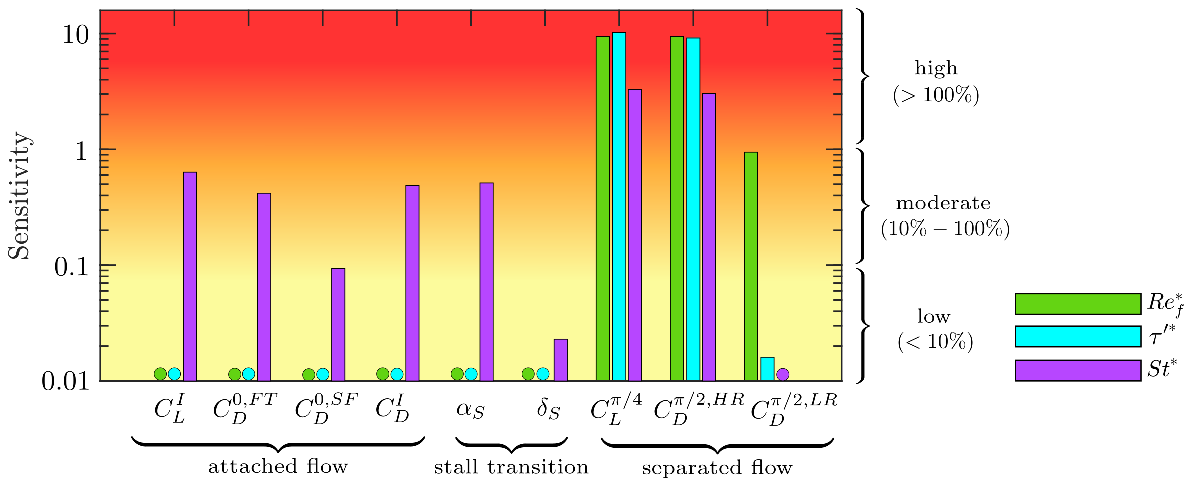}
   \caption{Sensitivity of the dynamical outputs to the constant parameters in the aerodynamic model. The relevant quantities include the critical $\Rey_f^*$ (green), characteristic growth/decay timescale $\tau'^*$ (cyan), and terminal Strouhal number $St^*$ (purple). Here, the dimensionless mass and amplitude are fixed with $M=A=1$, and $\Rey_f=10^5\gg\Rey_f^*$ for the results pertaining to $St^*$ and $\tau'^*$. Sensitivity scores less than $1\%$ are not displayed and instead marked with dots.}
  \label{fig:sensitivity}
\end{figure}

The results of this analysis are displayed as the bar chart of Fig. \ref{fig:sensitivity} showing the sensitivities of $\Rey_f^*$ (green), $\tau'^*$ (cyan), and $St^*$ (purple) with respect to the 9 aerodynamic parameters. Here scores less than $0.1=10\%$ can be considered low and hence the output is robust to changes in the parameter, and scores above $1=100\%$ are high and signify a strong dependency. Moderate sensitivity occurs for intermediate values, and very low scores below $0.01=1\%$ are not displayed and instead marked with dots. Evidently, the quantities $\Rey_f^*$ and $\tau'^*$ display sensitivity only to the coefficients $C_L^{\pi/4}$, $C_D^{\pi/2,HR}$ and $C_D^{\pi/2,LR}$, which are the subset that govern the separated flow regime. This suggests that the aerodynamics associated with attached flow and stall do not play a role in the stationary-to-locomotion transition. In contrast, the Strouhal number $St^*$ is broadly sensitive to nearly all parameters at levels that are at least moderate. Thus, it seems that the mechanisms associated with attached flow, stall, and separated flow are all important determinants of $St^*$. These observations are explained by the equilibrium and stability analyses provided in the next section.

It should be noted that nearly all 9 parameters affect at least one output quantity at a moderate level of $10\%$ or greater. The exception is $\delta_S$, which defines the angular breadth of the stall transition and displays uniformly low sensitivities. Even this parameter is necessary to include in the model in order to mathematically specify the crossover between the flow regimes, but apparently the outputs are robust to its numerical value. Taken together, these results indicate that the model is minimal in the sense of containing no unnecessary parameters.

\section{Interpretations via equilibrium and stability analyses}

The tractability of the quasi-steady formulation allows for analytical insights into key features of the dynamics, and we first seek interpretations for the critical Reynolds number $\Rey_f^*$ for take-off from rest and the characteristic dimensional time scale $\tau$ to reach the terminal locomotion state. Our theory is based on the linear stability of the stationary state, for which the attack angle is high and we may safely assume the separated flow forms of the lift and drag coefficients. It proves adequate to focus on one stroke in the flapping cycle and assume steady vertical motion $v_y \cong \overline{v}_y = 4af$ which here is chosen to be the mean speed. Perturbation of the flight speed introduces the small parameter $\varepsilon = v_x/v_y \ll 1$ whose dynamics we seek through Eq. \ref{ODEsystem}. The terms involve $v = \sqrt{v^2_x+v^2_y} = v_y\sqrt{\varepsilon^2+1} \cong v_y$ and $\Rey = \rho \ell v/\mu \cong \rho \ell v_y/\mu \cong 4 \rho \ell a f/\mu = 4 \Rey_f$, where the symbol ``$\cong$'' is used wherever the steady motion approximation or linearization in $\varepsilon$ is invoked. Inputting the attack angle $\alpha \cong \pi/2-\varepsilon$ yields the coefficients of lift $C^S_L(\Rey,\alpha) = C^{\pi/4}_{L}\sin2\alpha \cong C^{\pi/4}_{L} \sin (\pi-2\varepsilon) \cong 2 C^{\pi/4}_{L} \varepsilon$ and drag $C^S_D(\Rey,\alpha) = (C^{\pi/2,HR}_D + C^{\pi/2,LR}_D / \Rey) \sin^2 \alpha \cong (C^{\pi/2,HR}_D + C^{\pi/2,LR}_D / 4\Rey_f) \sin^2 (\pi/2-\varepsilon) \cong C^{\pi/2,HR}_D + C^{\pi/2,LR}_D / 4\Rey_f$. Inserting all these forms into Eq. \ref{ODEsystem} and simplifying yields an expression for the linearized dynamics:
\begin{equation}\label{eq:stabilityanalysis}
    \begin{split}
        \dot{\varepsilon} = \tau^{-1} \varepsilon ~\Rightarrow~ \varepsilon(t) = \varepsilon_0 e^{t/\tau} \textrm{,}~ \tau^{-1} = \frac{1}{2} \frac{\rho \ell v_y}{m+m_{11}}\left[2 C^{\pi/4}_{L} - \left(C^{\pi/2,HR}_D + \frac{C^{\pi/2,LR}_D}{4\Rey_f}\right)\right].
    \end{split}
\end{equation}
Hence the perturbation is predicted to decrease or increase according to the sign of $\tau$, which is the associated exponential decay or growth timescale. Notably, $\tau<0$ for sufficiently small $\Rey_f$, indicating that the stationary state is stable, whereas $\tau>0$ for larger $\Rey_f$ and the wing takes off into forward flight. The critical value $\Rey^*_f$ marking the transition is determined by $\tau^{-1}(\Rey^*_f)=0$:
\begin{equation}\label{eq:criticalRef}
    \begin{split}
        Re^*_f = \frac{C^{\pi/2,LR}_D}{4(2 C^{\pi/4}_{L} - C^{\pi/2,HR}_D)} = 25,
    \end{split}
\end{equation}
where the computed result reflects the nominal set of model parameters. This matches the critical value identified in the numerical results presented in Sections 3 and 4.

These results interpret the take-off condition for forward flight as requiring sufficiently high Reynolds number such that the horizontal component of lift exceeds drag at high attack angles, thereby generating a forward directed force vector that destabilizes the stationary state. The expression Eq. \ref{eq:criticalRef} also explains why the critical Reynolds number is almost entirely independent of $M$ and $A$ per the results of Fig. \ref{fig:parameter_sweep}(c), and why it is sensitive to the three aerodynamic coefficients $C^{\pi/4}_{L}$, $C^{\pi/2,HR}_D$ and $C^{\pi/2,LR}_D$ of the separated flow regime and insensitive to the other model parameters per the results of Fig. \ref{fig:sensitivity}. Similarly, this analysis interprets the timescale $\tau$ via Eq. \ref{eq:stabilityanalysis} as arising from the inertial resistance to motion in response to the aerodynamic forces. The dimensionless form and its limit for $Re_f \gg Re^*_f$ are
\begin{equation}\label{eq:tauprime}
    \begin{split}
        (\tau')^{-1} = 2\frac{A}{M}\left[2 C^{\pi/4}_{L} - \left(C^{\pi/2,HR}_D + \frac{C^{\pi/2,LR}_D}{4Re_f}\right)\right] \rightarrow 2\frac{A}{M}\left(2 C^{\pi/4}_{L} - C^{\pi/2,HR}_D\right) = \frac{A}{2.5M}.
    \end{split}
\end{equation}
This explains why the rescaled quantity $A\tau'^*/M$ is nearly conserved as shown in Fig. \ref{fig:parameter_sweep}(d), and the predicted value of 2.5 corresponds well with the range of 1 to 3 seen in the numerical results. The above expression also explains the sensitivity of $\tau'^*$ to the separated flow parameters and insensitivity to all others per Fig. \ref{fig:sensitivity}. 

A similar attempt to account for the selected Strouhal number views the forward flight state as arising from a balance of horizontal forces in which the forward-inclined lift vector balances the rearward-pointing drag. We again assume constant $v_y \cong 4af$, but $v_x$ cannot be assumed to be small. Equilibrium requires that $C_L(\alpha,\Rey)\sin\alpha=C_D(\alpha,\Rey)\cos\alpha$ and hence $C_L(\alpha,\Rey)/C_D(\alpha,\Rey)=\cot\alpha$. This condition represents an implicit relation for the unknown $v_x$, which appears in the attack angle $\tan \alpha = v_y/v_x \cong 4af/v_x = 2St$ and Reynolds number $\Rey = \rho \ell \sqrt{v^2_x + v^2_y} / \mu$. The equilibrium solutions are those $\alpha$ marking the intersections shown in Fig. \ref{fig:CL_CD_full}(f). Further analytical progress is hindered because the dynamically selected values $\alpha = 18^\circ$ to $27^\circ$ fall in or near the stall range, necessitating consideration of the full forms of $C_L$ and $C_D$ including the stall parameters and those of the attached and separated flow regimes. Nonetheless, the relatively narrow range of $\alpha$ across all $\Rey > \Rey_f^*$ implies a similarly conserved Strouhal number $St^*\cong(\tan\alpha)/2 = 0.16$ to $0.21$, which aligns with the findings of Figs. \ref{fig:bifurcation}(e) and \ref{fig:parameter_sweep}(a). Further, that the associated attack angles span the flow regimes explains why $St^*$ is sensitive to the full array of model parameters per Fig. \ref{fig:sensitivity}. We also note that these complexities similarly hinder any analytical accounting for the value  $\Rey_f=18$ marking the lower boundary of the hysteresis window and which would presumably be determined by loss of equilibrium for the locomoting state.

\section{Discussion and conclusions}

This study extends the quasi-steady model (QSM) framework for intermediate-$\Rey$ aerodynamics to account for the emergent flight dynamics of a thin plate undergoing vertical heaving oscillations in a fluid. This idealized setting serves as a testbed that allows for comparison to previous results from experiments \citep{vandenberghe2004symmetry, ramananarivo2016flow} and direct numerical simulations \citep{alben2005coherent, lu2006dynamic}, and for which the behavior is governed by relatively few dimensionless parameters, namely the mass $M$, flapping amplitude $A$, and flapping Reynolds number $\Rey_f$. By formulating aerodynamic coefficients to include appropriate dependencies on the Reynolds number $\Rey$ and attack angle $\alpha$, we show that the model reproduces the hallmarks of flapping-based propulsion. In particular, it captures the well known transition from a stationary state to forward locomotion that occurs as a spontaneous symmetry breaking at a critical value of $\Rey^*_f$ \citep{vandenberghe2004symmetry, alben2005coherent}. Associated subtleties such as hysteresis and bistability are also reproduced. Additionally, for sufficiently large $\Rey_f$, the system robustly converges to a nearly constant Strouhal number and hence the flight speed is proportional to the flapping speed. Moreover, the attained value of $St^*=0.2$ is consistent with previous experimental and computational studies and generally with the flapping propulsion of a wide array of aerial and aquatic organisms \citep{nudds2004tuning, taylor2003flying, ramananarivo2016flow}. Our model also furnishes predictions for the timescale $\tau'$ associated with reaching a terminal flight state and contributes the scaling relation $\tau'^* \sim M/A$ for sufficiently large $\Rey_f$.

Our model employs nine constant parameters, the nominal values of which we have suggested by appealing to previous studies of the aerodynamic forces on thin plates in steady flow \citep{li2022centre, bhati2018role, ehrenstein2013skin, ananda2015measured, gutierrez2021lift}. The general phenomena and scaling trends are robust without any particular tuning of the parameters, while the numerical values of output quantities show different degrees of sensitivity. Our characterizations focus on the three main dynamical outputs: the transitional Reynolds number $\Rey^*_f$, the dimensionless timescale $\tau'^*$ for sufficiently large values of $\Rey_f$, and the emergent Strouhal number $St^*$ in the same limit. The values of $\Rey^*_f$ and $\tau'^*$ depend on the aerodynamic coefficients relevant to separated flow at high attack angles, which is explained by a linear stability analysis of the stationary state. In contrast, the dynamically selected $St^*$ is sensitive to a broader array of parameters, which is consistent with an equilibrium analysis of the forward flight state showing that the characteristic angle of attack is near the stall transition for which the full slate of aerodynamic coefficients are important.

Our model uses quasi-steady aerodynamics based on steady lift and drag to address dynamical flight behaviors that previous studies have linked to unsteady vortex-based mechanisms \citep{vandenberghe2004symmetry, alben2005coherent}. For example, the selected $St^*$ characterizing steady forward flight has been connected to specific modes of vortex shedding, whereas our model sees it as a condition of force balance whose robustness reflects the relative invariance of the lift and drag coefficients across $\Rey = 10^2$ to $10^5$. Similarly, the take-off effect at a critical $\Rey^*_f$ has been attributed to wing-vortex interactions, whereas our model frames it as an instability induced by a transition from drag- to lift-dominated aerodynamics as $\Rey$ is increased. The unsteady and quasi-steady views might be reconciled by seeing their apparent differences as largely interpretative and semantic, with one emphasizing the causal flows and the other the effective forces. It may be that the model gives fair account of the stroke-averaged forces, though likely not the details of their variations in time. This hypothesis should be systematically tested in future work, for example by comparing to experimental force/torque measurements or direct numerical simulations under matched conditions.

The dynamical model provided here is directly applicable to 2D flight of a thin, rigid and symmetric wing undergoing simple heaving. It will be useful to test the model for a broader set of flapping kinematics, for example pitching oscillations or combined pitching and heaving motions. Such work may start by inserting the force coefficients derived here into the more general model of \citet{Pomerenk_Ristroph_2025} that includes rotational dynamics. Extensions to non-slender and/or asymmetric profile shapes such as foils would require appropriate modifications to the force coefficients informed by experiments or simulations \citep{ wang2000vortex, lu2008three, shyy2010recent, jardin2012three}. Flexible structures that involve significant dynamical deformations within a stroke seem beyond the reach of such a QSM framework, and there are open questions about if and how three-dimensional effects might be addressed. Despite the inherent limitations of quasi-steady aerodynamic modeling, the encouraging results presented here motivate future efforts to bring additional aspects of intermediate-$\Rey$ motion and locomotion under their purview.

\textit{Acknowledgments.} We thank S. Childress, M. Shelley, Z. J. Wang and J. Zhang for useful discussions.

\textit{Funding.} This work was supported by the U.S. National Science Foundation through the award DMS-1847955.

\textit{Declaration of interests.} The authors report no conflict of interest.

\textit{Author ORCID.} O. Pomerenk, https://orcid.org/0000-0002-0481-3952; L. Ristroph, https://orcid.org/0000-0001-9358-0689.

\bibliographystyle{jfm}
\bibliography{jfm}

\end{document}